\documentclass[a4paper,twoside]{article}

\usepackage{epsfig}
\usepackage{subcaption}
\usepackage{calc}
\usepackage{amssymb}
\usepackage{amstext}
\usepackage{amsmath}
\usepackage{amsthm}
\usepackage{multicol}
\usepackage{pslatex}
\usepackage{apalike}
\usepackage{algorithm2e}
\usepackage{url} 
\usepackage[bottom]{footmisc}
\usepackage{SCITEPRESS}     
\usepackage{pgfplots}
\usepackage{graphicx}
\begin{document}

\title{Real-Time Kinematic Positioning and Optical See-Through Head-Mounted Display for Outdoor Tracking: Hybrid System and Preliminary Assessment}


\author{\authorname{Muhannad ISMAEL\orcidAuthor{0000-0003-4274-9311} and Maël Cornil \orcidAuthor{0000-0002-6325-5997} }
\affiliation{Luxembourg Institute of Science and Technology (LIST)\\Esch-sur-Alzette, Luxembourg }
\email{\{muhannad.ismael, mael.cornil\}@list.lu}
}

\abstract{This paper presents an outdoor tracking system using Real-Time Kinematic (RTK) positioning and Optical See-Through Head Mounted Display(s) (OST-HMD(s)) in urban areas where the accurate tracking of objects is critical and where displaying occluded information is important for safety reasons. The approach presented here replaces 2D screens/tablets and offers distinct advantages,  particularly in scenarios demanding hands-free operation. The integration of  RTK, which provides centimeter-level accuracy of tracked objects, with OST-HMD represents a promising solution for outdoor applications. This paper provides valuable insights into leveraging the combined potential of RTK and OST-HMD for outdoor tracking tasks from the perspectives of systems integration, performance optimization, and usability. The main contributions of this paper are: \textbf{1)} a system for seamlessly merging RTK systems with OST-HMD to enable relatively precise and intuitive outdoor tracking, \textbf{2)} an approach to determine a global location to achieve the position relative to the world, \textbf{3)} an approach referred to as 'semi-dynamic' for system assessment. Moreover, we offer insights into several relevant future research topics aimed at improving the OST-HMD and RTK hybrid system for outdoor tracking.}

\onecolumn \maketitle \normalsize \setcounter{footnote}{0} \vfill

\section{\uppercase{Introduction}}
The primary motivation of this work is to explore the integration of  OST-HMD  and RTK systems for outdoor tracking, particularly in the context of managing CBRN (Chemical, Biological, Radiological, and Nuclear) incidents. Our focus here is on radiological incidents. Incident management involves numerous first responder organizations, as well as potentially the military and other agencies. Those involved in responding to such incidents require accurate, real-time information regarding the risks present and the positioning and utilization of assets such as Unmanned Aerial and Ground Vehicles (UAVs, UGVs) to detect and identify sources of contamination.

Leveraging OST-HMD with RTK systems for real-time outdoor tracking could significantly enhance situational awareness during radiological incidents. By providing first responders with the ability to perceive, comprehend, and plan appropriate courses of action based on real-time accurate information, this technology can help mitigate risks and manage incidents more effectively. However, radiological incidents can occur in diverse environments, including densely populated urban areas, under various lighting conditions, and in different weather conditions.


Most contemporary technologies designed for outdoor object tracking with high precision primarily rely on image inputs. The utilization of  machine learning approaches, such as YOLOv5 \cite{DBLP}, facilitates the detection and tracking of various objects. Nevertheless, these methods encounter challenges when objects are concealed by obstacles, in adverse weather conditions, or during nighttime. Consequently, their effectiveness diminishes under such conditions. In response to these limitations, alternative methods incorporating GPS signals have been employed, offering estimations with limited precision \cite{stranner2019high}. However, certain applications demand even greater accuracy, prompting the utilization of RTK systems to achieve enhanced tracking precision. 

To visualize the tracking information obtained from  GPS or RTK systems, conventional approaches employ tablets or 2D screens. Advancements in OST-HMDs technology have paved the way for novel applications that capitalize on the benefits of observing tracked objects through OST-HMDs. These devices are based on the half-silvered mirrors technique to merge the view of virtual and real objects. The advantage of this technique is the ability to directly view the real world as is and not via a computer rendering as is the case with \textbf{V}ideo \textbf{S}ee \textbf{T}hrough HMDs (VST-HMDs). This avoids problems with lag, and often reduces other ergonomic issues associated with VSTs, such as discomfort and heat.

Despite their impressive capabilities for blending digital content with the real world, OST-HMDs  are not recommended for outdoor scenarios. Most OST-HMDs rely on depth-sensing cameras to map the user's environment and interact with virtual objects. While they can work outdoors, the performance of depth sensing may degrade in bright sunlight or on highly reflective surfaces, leading to less precise spatial mapping and interaction. Moreover in bright sunlight, the display may appear less vibrant, and the virtual objects content may be less visible compared to indoor environments. \\
However, OST-HMDs offer several advantages over traditional 2D visualization: 1) spatial awareness supported by OST-HMDs allows to tracked objects in their actual surroundings, making it easier to comprehend their positions and movements; 2) users can interact with the tracked objects in a hands-free manner. This is particularly beneficial in scenarios where users need to focus on tasks or have limited physical mobility; 3)  OST-HMDs can offer intuitive navigation assistance by overlaying visual cues or directions onto the real-world environment. This can be particularly useful for guiding users to specific tracked objects or locations. 

\begin{figure*}[ht]
    \centering
    \includegraphics[scale=0.3]{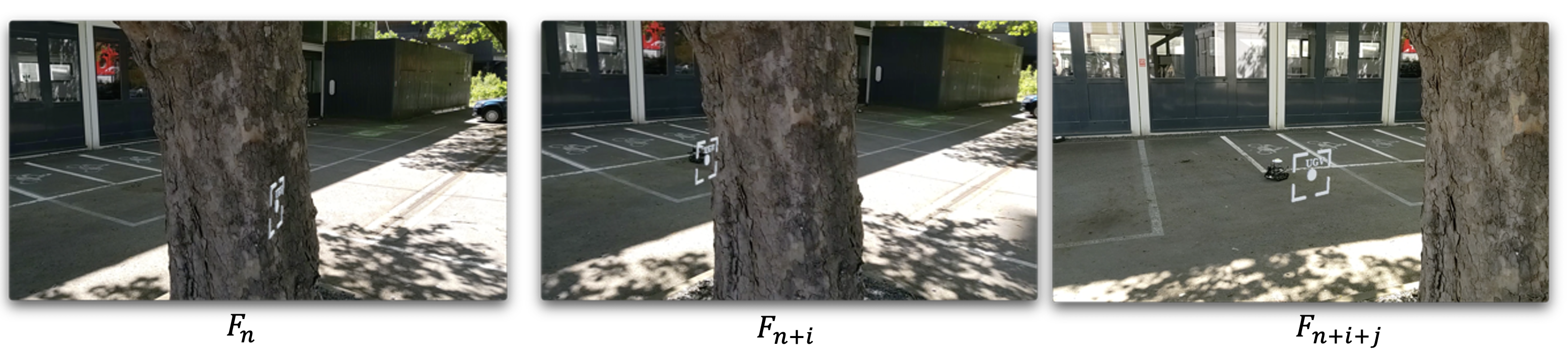}
    \caption{ Tracking UGV using RTK systems. $F_{n}$, $F_{n+i}$ and $F_{n+i+j}$ are captured frames from Microsoft HoloLens v2. The white virtual rectangle refers to the RTK information derived from RTK rover system located on UGV.}
     \label{tracking_robot_car}
\end{figure*}

Despite the fact that OST-HMDs are not fully adapted for outdoor use, these advantages have attracted numerous researchers \cite{8798315,satheesan2024real,oskiper2012multi} to analyze potential scenarios for their application such as tracking real object in outdoor environment. In this paper, we suggest a system that integrates the RTK system with OST-HMD for outdoor scenario. We integrated an RTK system and a Raspberry Pi into a UGV, connected to the proposed web-based server. Moreover, a semi-dynamic approach is proposed to evaluate the system and illustrated in Section \ref{experiment_and_discussion}. The paper presents preliminary results, shedding light on the potential of this integrated system, while also highlighting the myriad challenges associated with its implementation. Hence, in this paper, we aim to address the following research questions: 
\begin{itemize}
    \item RQ1: how can UGVs be effectively visualized using OST-HMD when obstacles obstruct the view of objects as illustrated in Figure \ref{tracking_robot_car}?
    \item RQ2: how can data from RTK systems be seamlessly integrated into OST-HMD?
    \item RQ3: how well does the semi-dynamic approach adapt to the challenges of real-world UGV tracking evaluation compared to the static and dynamic methods?
    \item RQ4: what are the key research directions for optimizing the integration of OST-HMD with RTK system for outdoor traking?
\end{itemize}
\section{\uppercase{Background}}
\label{background}
We review the related works in Sections \ref{subsectionRTK_systems_with_OST-HMDs}, \ref{subsectionGlobalLocation}, and \ref{subsectionAccuracyEvaluation}, which correspond to our three main contributions: 1) a system that seamlessly integrates RTK technology with OST-HMD to enable accurate and intuitive outdoor tracking, 2) a method for determining global location to establish position relative to the world, and 3) a 'semi-dynamic' approach used for system evaluation.


\subsection{ RTK system with OST-HMD}
\label{subsectionRTK_systems_with_OST-HMDs}
GPS is composed of 24 satellites that orbit at medium altitudes around the Earth. Although it provides 3D positions, time, and velocity data, several drawbacks limit GPS systems to meter accuracy \cite{morales2007dgps}. RTK systems overcome these limitations by using correction signals generated from an additional base station, achieving centimeter accuracy \cite{gan2007implement}. However, numerous studies \cite{niu2020continuous} confirm that the performance of RTK systems is closely tied to the geographical context in which they operate. Specifically, the accuracy of these systems in urban areas is notably lower compared to open-sky environments \cite{de2023systematic}. This discrepancy can be attributed to various factors such as signal obstructions, multi-path interference, and the density of structures in urban settings. Consequently, users should consider these environmental factors when assessing the reliability and precision of RTK-based positioning in different geographical regions. Moreover, we can delineate two distinct categories within the realm of RTK systems: 1) Large-size RTK systems \cite{klein07parallel} offer heightened precision and extended baselines between the base station and rover, albeit at a higher cost; 2) conversely, small-size RTK devices \cite{de2023systematic} prioritize portability, making them more suitable for on-the-go applications. For example, a study proposed by \cite{de2023systematic} evaluated a smartphone RTK device, which is small-sized, from the REDCatch company\footnote{https://www.redcatch.at}. The results show that the RTK device can achieve 1 cm accuracy when used in open-sky areas. In contrast, its accuracy dramatically reduces in the proximity of buildings and obstacles.

In this paper, a preliminary experimental test was conducted in densely populated urban areas to simulate a realistic scenario for potential system deployment in CBRN context management. Furthermore, a large-size RTK system is proposed for use to achieve high accuracy without requiring the portability of the antenna by the person wearing the HMD device. This is made possible by the hybrid system combining RTK and vSLAM, as detailed in the next Section.  

In the current state of the art, the implementation of Augmented Reality (AR) with handheld devices (e.g., tablet, smartphones, etc.) for outdoor environments using GPS or RTK systems has been addressed in numerous studies \cite{singh2022augmented,schall2009global,stranner2019high,niu2020continuous}. However, OST-HMDs such as the HoloLens and VST-HMDs such as the Apple Vision Pro and Meta Quest have not yet been fully adapted for outdoor use. Despite this, researchers are actively endeavoring to assess the suitability of HMDs for outdoor applications \cite{8798315,satheesan2024real}. 

OST-HMDs (e.g., the Microsoft HoloLens version 2) utilize visual Simultaneous Localization and Mapping (vSLAM) techniques to generate a map of the surroundings and determine their own location within it \cite{ungureanu2020hololens,zari2023magic}. vSLAM is a specific subset of SLAM that relies primarily on visual information from cameras for localization and mapping. However, these algorithms depend on the inputs from the depth sensing cameras and are primarily designed for indoor use. RTK systems could potentially overcome these limitations. However the objective in this paper is not to improve the vSLAM algorithm of OST-HMD for outdoor environments, our main contribution is to propose a system to integrate the information derived from RTK and track this information using vSLAM algorithm.  

Hence, in this Section, our focus is on integrated HMD with GPS and RTK systems. Consequently, we categorize these studies into two groups: those based on GPS/RTK systems and hybrid-based approaches, as outlined below:

\subsubsection {GPS/RTK}
One of the pioneering studies integrating GPS with HMDs was proposed by \cite{Swan2003PerceptualAE}, where they introduced an AR application aimed at enhancing urban operations. The application is based on Battlefield Augmented Reality System (BARS) which  was mainly composed of Sony Glasstron LDI-D100B, Ashetech GG24-Surveyor, InterSense, InertiaCube2, etc. Work by \cite{dahne2002archeoguide} suggests ARCHEOGUIDE (Augmented Reality-based Cultural Heritage On Site GUIDE) which aimed to develop an interactive methods for accessing cultural heritage information via HMD and using GPS information. A study by  \cite{behzadan2005visualization} 
suggests to provide architects the ability to view virtual models of construction in an urban environment. A system called "Mobile Augmented Reality System" MARS  was proposed by \cite{hollerer1999exploring} that allows visualising  outdoor spatial information. All these studies suffer from inaccurate outdoor location since the tracking methods are only based on GPS/RTK receivers.\\
Work by \cite{roberts2002use}  highlights the benefits of merging AR and RTK systems to see underground features. These features could be major geological structures, gas or water pip-work or zones of contaminated land. However, the authors don't describe  how to merge AR and RTK systems.   The system is fixed in one position and overlay virtual and real objects form a fixed position and orientation. Moreover, it was one of the early works in this field, and HMDs devices were less advanced in comparison to their current state. \\
Hence, AR visualization via HMDs  based only on GPS/RTK information is not accurate enough. Therefore, many current approaches focus on developing hybrid systems that integrate data from multiple sensors.

\subsubsection{Hybrid}
Numerous studies suggest hybrid GPS/RTK systems with  information derived from other sensors  from handheld devices  \cite{burkard2020user,schall2009global}.  However, in the state of the art, hybrid systems with HMDs are less explored for outdoor applications.  A system with a Monocular Camera, Inertial Measurement Unit (IMU) and GPS unit was proposed by \cite{oskiper2012multi} to track the camera motion in 6 degrees of freedom in arbitrary indoor and outdoor scenes. They used an HMD with an IMU and GPS. They compared their results against  GPS-only solution and found that it had an advantage over a GPS-only approach.

According to \cite{satheesan2024real}, a proof-of-concept system was developed for operators using HMDs in outdoor conditions. The system aimed to overlay an offline point cloud map of the parking area onto real-world objects visible through the HMD. The Varjo XR-3\footnote{https://varjo.com/products/varjo-xr-3/} and NovAtel CPT-7\footnote{https://novatel.com/}, which provide global positioning, were used in the proposed system. The results showed that the point cloud partially matched real-world objects for 66.7\% of the total duration, with observed discrepancies in depth alignment between the point cloud and the real-world objects.

 Recently, a hybrid system using RTK with vSLAM was developed to obtain high-precision outdoor tracking using a  Microsoft HoloLens  \cite{8798315}. The proposition is close to our suggestion in this paper. However, a few significant limitations exist in the previous system. To begin with, no formal evaluation or performance benchmarking of the system was provided, leaving questions about its practical accuracy and reliability in real-world conditions. Additionally, the system requires the user to attach an external antenna to the head-mounted display (HMD), potentially hindering mobility and comfort. Furthermore, the proposed system updates the user's position using RTK information for each frame, which may not be an optimal approach; frequent RTK updates may introduce latency, as RTK signals can be subject to delays depending on environmental factors such as signal interference or obstructions. In contrast, our system uses RTK to determine an initial reference position, after which the vSLAM algorithm takes over to track the user's position. This method significantly reduces the need for constant RTK updates combining RTK’s high initial accuracy with vSLAM’s real-time adaptability. Moreover, this approach of combining RTK and vSLAM leads to a more portable solution by eliminating the need for external antennas while maintaining tracking accuracy. We also provide a preliminary assessment of our system’s performance including accuracy as described in details in Section \ref{experiment_and_discussion}.

\subsection{Global location}
\label{subsectionGlobalLocation}
vSLAM can track and create a local map of the areas traversed by the observing device. However, it cannot provide a global location relative to the world. One way to achieve such a global location is by matching the observed data to another dataset containing geotagged information. For example, a study by \cite{zhang2006image} used pre-existing images tagged with GPS coordinates. An image from the camera was then used to find matches in the image database. Another approach utilizes a 2.5D technique \cite{arth2015instant}, where pre-established untextured building structures with accurate GPS locations are combined with SLAM observations. The end device employs a coarse GPS location to identify matches among the existing 2.5D structures stored on the device. To achieve an accurate pose, the geometry of the stored 2D data is compared with the models available on the device. Like the previous suggestion \cite{zhang2006image}, this approach also requires initial preparation, specifically the work needed to create the 2.5D models for the desired areas.

Instead of relying on images or structural shapes, global positioning can be based on an available 3D map file saved on the device \cite{lim2012real}. This file can contain GPS positions and the global location of the device itself without the need to rely on online solutions. 

In the field of deep learning and positioning, work by \cite{rao2017mobile} used an existing object detection method called SSD (Single Shot Detector) to detect objects. To be able to use SSD on mobile devices, they modified it to be less computationally heavy. They provided large image sets to the learning algorithm and processed them with a powerful computer. A mobile device was then utilized, which provided rough GPS data, together with data from IMU (inertial measurement unit) sensors and a magnetometer, along with a camera. As a result, almost real-time object recognition was achieved with a frame rate of two frames per second. \\
In this paper, we propose a novel approach for determining a global position relative to the world, based on a reference location determined using RTK system. Initially, the person wearing OST-HMD stands at the same location as the UGV, which is equipped with RTK systems. This location is used as the reference point. The orientation is also calibrated to align the OST-HMD' coordinate frame with the world coordinate frame. Subsequently, the UGV's position is tracked on the OST-HMD using this reference point, as described in detail in Section \ref{AR_Application} 

\subsection{Accuracy Evaluation}
\label{subsectionAccuracyEvaluation}
The accuracy of a system containing RTK or GPS can be evaluated in static or dynamic scenarios. The RTK or GPS systems remain fixed at a specific location  \cite{wisniewski2013evaluation,safrel2018accuracy} in static conditions, while it keeps changing its physical location in dynamic conditions \cite{kluga2014state,tomaszewski2020assessment}. Work by \cite{stranner2019high} compared a low-cost RTK device combined with IMU, altimeter, and camera with a highly accurate RTK receiver in static conditions. A collaborative virtual environment was developed by \cite{singh2022augmented}, allowing users to interact through a handheld device tracked using a vSLAM approach in conjunction with traditional GPS. The authors compared their tracking solution under static conditions to conventional GPS and achieved a more stable and smoother trajectory than that generated by standard GPS.

The accuracy of mapping virtual objects onto the real world was evaluated using the Varjo XR-3 in a dynamic method by \cite{satheesan2024real}. They recorded a video of the trajectory captured by the Varjo XR-3 and manually annotated the frames to assess the alignment between the real and virtual objects.  

We propose another approach called semi-dynamic, where the UGV is captured at specific locations while completing the trajectory. In this paper, one of our key contributions is a novel approach to evaluate the accuracy of UGV tracking using a semi-dynamic method, as detailed in Section \ref{experiment_and_discussion}. This approach focuses on the methodology rather than the results, as it is well known that RTK provides higher accuracy than GPS.

\section{\uppercase{System}}
\subsection{scenario}
\label{sect_scenario}
As highlighted in the introduction, the focus of this work revolves around enhancing the management of CBRN incidents through the visualization of tracked a UGV. First responders, including firefighters, military personnel, and other emergency teams, often face critical situations where they must intervene to neutralize, for example, a source of radiation in an urban area. In these scenarios, the initial step involves deploying a UGV equipped with a specialized gamma camera \cite{gal2001cartogam}. This camera is designed to detect and pinpoint the exact location of the radiation source, enabling the team to address the threat effectively.

The key advantage of utilizing a UGV lies in its ability to perform reconnaissance and intervention tasks without putting human lives at risk. By keeping first responders at a safe distance, the UGV minimizes their exposure to dangerous levels of radiation and other associated hazards. This approach not only enhances the safety of emergency personnel but also improves the efficiency and precision of the intervention.

To further augment situational awareness and operational effectiveness, we propose that first responders wear OST-HMD. These can overlay critical information directly into the responders' field of vision, including real-time data on the UGV's position, radiation levels, and other vital metrics. This integrated system ensures that first responders have immediate access to comprehensive information, facilitating informed decision-making and coordinated actions during the intervention.\\ 
Hence, we hypothesize that by combining the capabilities of UGV with OST-HMD, we can significantly enhance the safety, accuracy, and efficiency of radiation neutralization efforts in urban environments. Consequently, in this investigation, we propose a system to visualize tracked UGVs in outdoor environments via OST-HMD. This system addresses the initial research question \textit{RQ1: how can UGVs be effectively visualized using OST-HMD when obstacles obstruct the view of object?} To answer this question, the system includes three main components, which are detailed in the following Sections: 

\begin{itemize}
\item The server application receives the positions of the UGV and transmits this data to the OST-HMD.
\item Sensors in our scenario are UGVs equipped with an RTK system.  This system is composed of an RTK rover and an RTK station. The position of the RTK rover is corrected using the RTK station, as will be illustrated later in Section \ref{sect_sensors}. The choice of the RTK system, as mentioned in the introduction, offers significant advantages over image-based tracking in scenarios involving obstacles, adverse weather, or nighttime conditions.
\item The AR application deployed on OST-HMD serves as an interface for visualizing information provided by the RTK rover.

\end{itemize}


\subsection {Assess system requirements}
\begin{figure}[ht]
     \centering
        \includegraphics[scale=0.3]{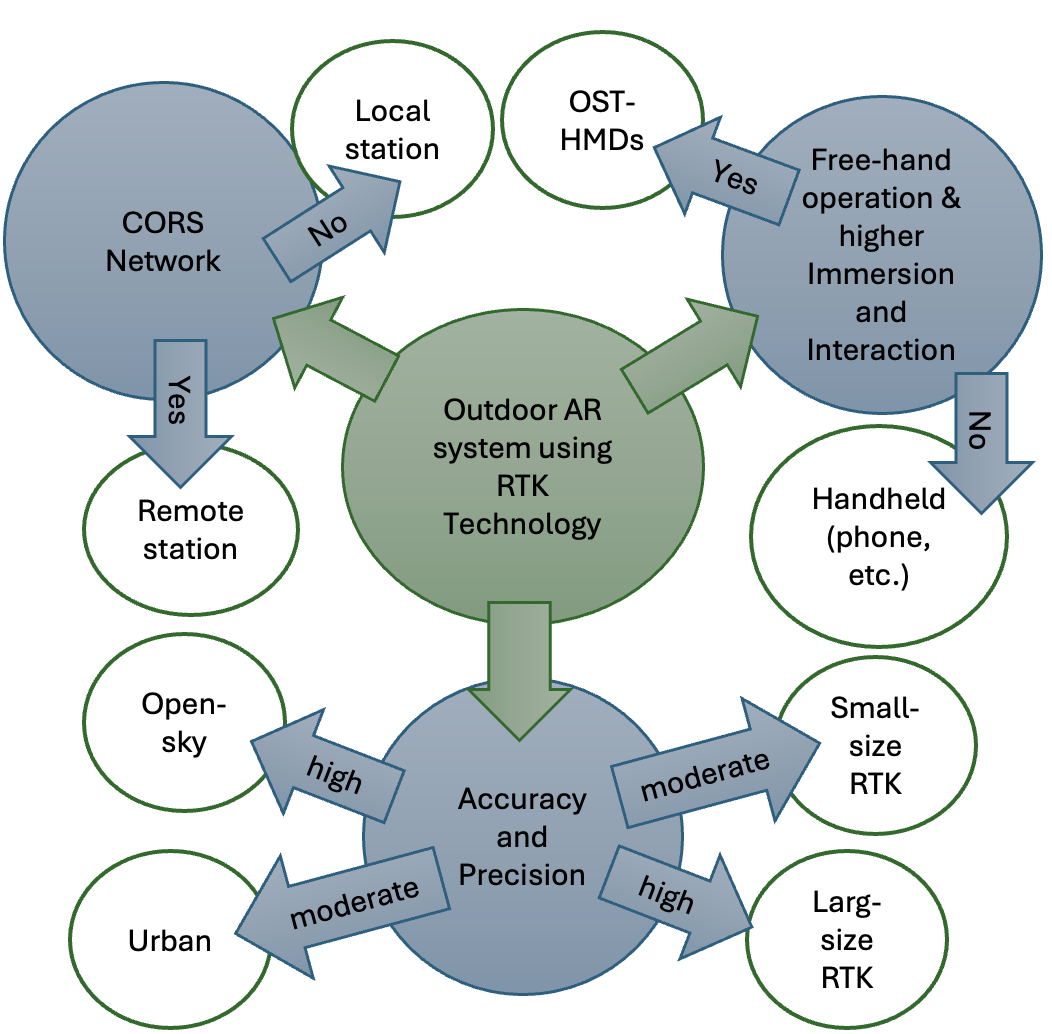}
	
     \caption{ System requirements for outdoor AR system using RTK technology. }
        \label{fig:diagram_system_analysis}
\end{figure}

\begin{figure}[ht]
     \centering
        \includegraphics[scale=0.6]{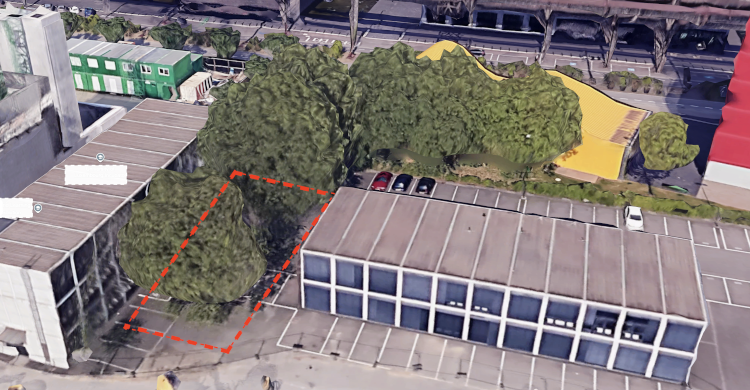}
	
     \caption{  Picture captured from Google Earth. The area highlighted in red is where the experimental test is conducted }
        \label{fig:env}
\end{figure}

In order to reach the decision to use RTK we conducted a detailed requirements analysis which is outlined in  Figure \ref{fig:diagram_system_analysis}. In CBRN management, free-hand operation for first responders is essential. For this reason, OST-HMDs were selected rather than  handheld AR devices. A Continuously Operating Reference Station (CORS) network consists of a series of fixed reference stations that continuously collect GNSS data. In such a system, the rover corrects its position often via an internet connection through a remote station provided by the CORS network. Some countries around the world provide CORS networks. In our case, the system was not supported by CORS; hence a local station was chosen. Furthermore, we opted for large-size RTK systems to achieve the highest possible positional accuracy. CBRN incidents often occur in urban areas, so we evaluated the system in these areas (see Figure \ref{fig:env}). Despite the fact that the RTK system provides less accuracy and precision compared to open-sky conditions.

\subsection{Server application}
\label{server_application}
We utilized Tomcat, a web hosting service built around the Java programming language. It offers a REST API as well as socket connections. This server is hosted on a web-based platform known as (Anonymous web server). The server receives the message from sensors and broadcasts to the OST-HMD.  Since real-time communication is required, raw socket connections are used. This offers lower latency compared to HTTPS due to reduced protocol overhead and encryption. 

\begin{figure}[ht]
    \centering
       \includegraphics[scale=0.25]{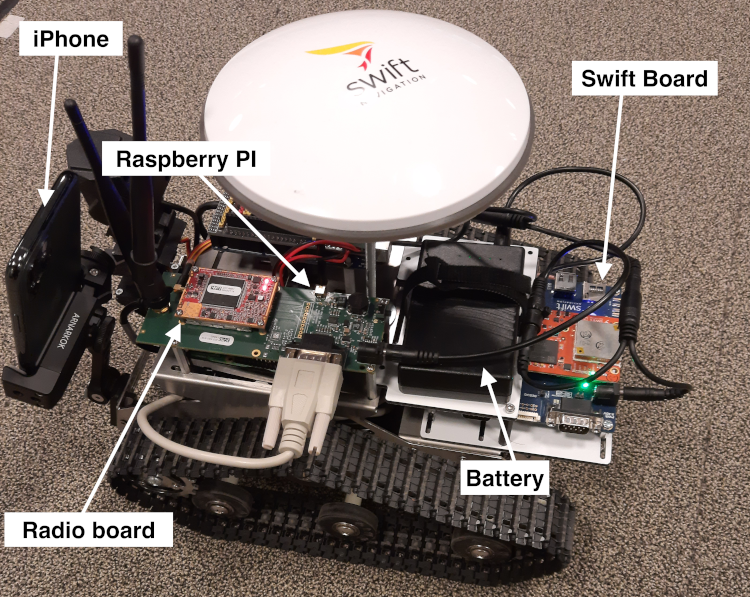}
    \caption{  UGV consists of an antenna, Swift Navigation Piksi board, Radio board, battery, phone holder, iPhone 11 Pro, and Raspberry Pi, which is hidden by a radio board that receives the correct position of the UGV using information derived from the RTK station }
       \label{fig:robotCar}
\end{figure}
\begin{figure}[ht]
     \centering
        \includegraphics[scale=0.4]{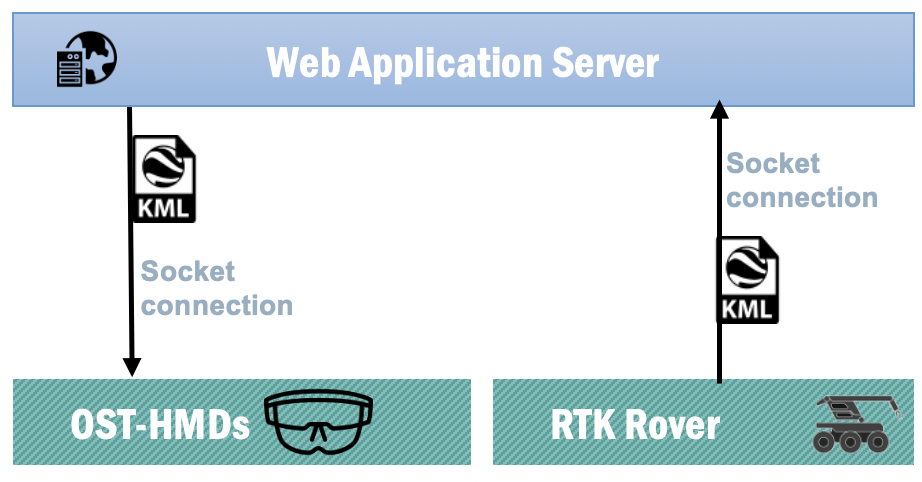}
	
     \caption{ General schema  of the proposed system}
        \label{fig:schema}
\end{figure}

\subsection{Sensors} \label{sect_sensors}
Sensors such as UAVs, UGVs, or other types can be used in CBRN management. In the scenario presented in Section \ref{sect_scenario}, the sensor specifically refers to one mounted on a UGV \footnote{https://www.xiaorgeek.net}.  This latter operates through a specialized application that communicates via a dedicated WiFi network. This application serves as the control interface, enabling users to remotely manage and command the UGV's movements. We customized the UGV to fit our requirements, incorporating components such as an RTK rover (an antenna, Swift Navigation Piksi board \footnote{https://www.swiftnav.com}, Radio board), phone holder, iPhone 11 Pro and Raspberry Pi on the UGV (see Figure \ref{fig:robotCar}). Moreover,  an RTK station (see Figure \ref{system}) is used to transmit GNSS correction data over the radio link to the RTK rover.  The RTK station position known as the "surveyed position" is determined manually or automatically. The surveyed position will be used to correct the position derived from RTK rover. In our case, the surveyed position was not available, therefore, we used automatic surveyed position which is generated using average of the last 1000 Single Point Positioning (SPP) position solutions. Furthermore, RTK rover interfaces with the Raspberry Pi, which facilitates the transmission of the UGV's position data to a web server known as (Anonymous web server). Moreover, an iOS application is developed and deployed on iPhone 11 Pro. The application provides GPS capabilities for location tracking and navigation using the CLLocationManager class. This information is then sent to the web server, similar to the information sent from the Raspberry Pi. 
 Utilizing location information derived from the iPhone provides us with the opportunity to compare it with RTK location information, as illustrated in Section \ref{experiment_and_discussion}.
\subsection{AR Application}
\label{AR_Application}
In this investigation, HoloLens version 2 is used. An AR application is developed using Unity3D. It's connected with UGV via the server application (see Figure \ref{fig:schema}).
Moreover, we propose to utilize sun protection filter \footnote{\url{https://www.realsim.info/en-gb/hololens-2-sonnenschutzfolie}} to reduce some of the effects of  bright sunlight on the outdoor experience. The HoloLens app is based on vSLAM to generate a map of the surroundings and find its own location within it. Our main contribution in this application is to respond to the \textit{RQ2: how can data from RTK systems be seamlessly integrated into OST-HMD?} To answer this question, the application contains two main functionalities: 1) calibration to obtain a global location, and 2) location update, as described below:
 \subsubsection{Calibration}
Global location consists of finding the correct position and orientation of the person wearing the HoloLens relative to the world. \\
\textbf{User's position: } 
At the start, the person wearing the HoloLens stands in the same location as the UGV. This location is considered as the reference point.  Hence, the position $P_{ref}^{world}$ derived from the RTK rover is the same as the position of the HoloLens device $P_{ref}^{HoloLens}$ , but in different coordinate frames. The first is in the World coordinate frame, and the second is in the HoloLens coordinate frame. The two  positions $P_{ref}^{World}$, $P_{ref}^{HoloLens}$ are saved to calculate the updated positions.  \\ 
\textbf{User's orientation: } The bearing angle indicates the angle between the reference position $P_{ref}^{World}$ and the UGV's position $P_{}^{World}$ relative to the north direction. This angle is crucial for navigation and positioning tasks. If the OST-HMD coordinate frame is aligned with the World coordinate frame, meaning that the negative z-axis of the HoloLens aligns with the north direction, then the bearing angles calculated in the  World coordinate frame will match those in the HoloLens coordinate frame. This alignment ensures consistency in directional references across both systems. To achieve this alignment, at the beginning, the user's head direction, while wearing the OST-HMD, should face the north direction. This initial orientation aligns the user's perspective with the  World coordinate frame, facilitating accurate bearing angle measurements and consistent spatial orientation
\subsubsection {Update position} 
Keyhole Markup Language (KML) messages are composed of the coordinates of the UGV in World coordinate frame are transmitted to the server application via the detected Wi-Fi network. The server relays this data to the AR application, which then displays the relevant information on OST-HMD. As mentioned previously, sockets are employed to enable multiple simultaneous messages between the various system components. Therefore, to compute the position of UGV in the HoloLens coordinate frame, we follow these steps
\begin{itemize}
    \item  Computing the distance $\delta$ between  UGV's position $P_{}^{World}$ and reference position $P_{ref}^{World}$  which refers to the previously saved reference position from the calibration step 
    
    \item Calculating the bearing angles between the reference position $P_{ref}^{World}$ and the UGV's position $P_{ref}^{World}$  in the World coordinate frame. This angle will be the same in the HoloLens coordinate frame thanks to the calibration step, which aligns the negative z-axis of the HoloLens with the north direction
\end{itemize}
Hence, the position  $P_v^{HoloLens}$  of the virtual object corresponding to the position of UGV in HoloLens coordinate frame is computed as follows:
\begin{equation}
    P_{v}^{HoloLens} = P_{ref}^{HoloLens} + P_{\delta}^{HoloLens}
\end{equation}
\begin{equation}
P_{\delta}^{HoloLens}  = 
\begin{pmatrix}
\delta \cdot \cos(\beta) \\
\delta \cdot \sin(\beta) \\
0
\end{pmatrix}
\end{equation}
Where $P_{\delta}^{hololens}$ refers to the  position after the rotate of point $( \delta,0,0)^{T}$ by the  bearing angle $\beta$  which indicates the angle between the reference and UGV positions according to north direction. Knowing that the negative z axes of HoloLens coordinates system is aligned with north direction via calibration steps as mentioned previously.  Hence, the AR application transforms the RTK coordinates of UGV derived from KML file into the  HoloLens coordinate frame using the reference position  $P_{ref}^{HoloLens}$. Therefore, the proposed approach isn't required to update the position of the user wearing the HoloLens in each frame, as in \cite{8798315}, thus avoiding noisy information and providing more stable results.
\begin{figure}[ht]
     \centering
        \includegraphics[scale=0.25]{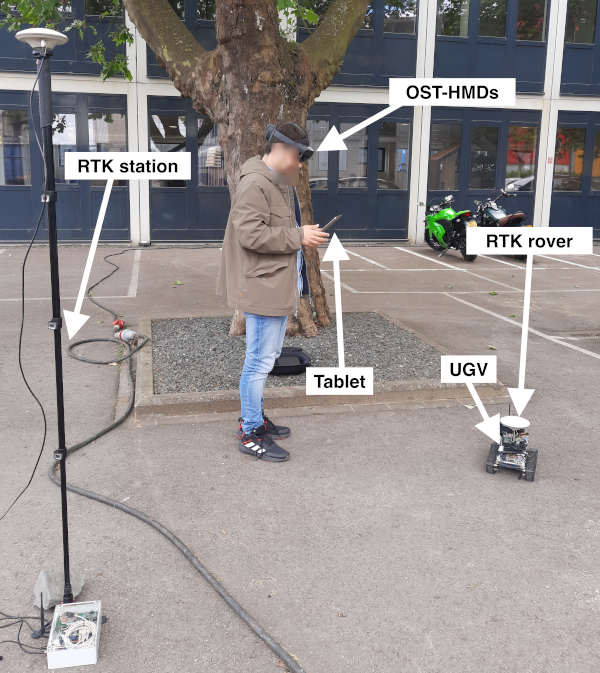}
     \caption{ A prototype consisting of a UGV, RTK rover (including an antenna, Swift Navigation Piksi board, and Radio board), RTK station, tablet with a specific application enabling users to remotely command the UGV's movements, and OST-HMD such as the HoloLens v2.}
        \label{system}
\end{figure}
\section{ \uppercase{Results}}
\label{experiment_and_discussion}
\begin{figure}[ht]
     \centering
         \includegraphics[scale=0.16]{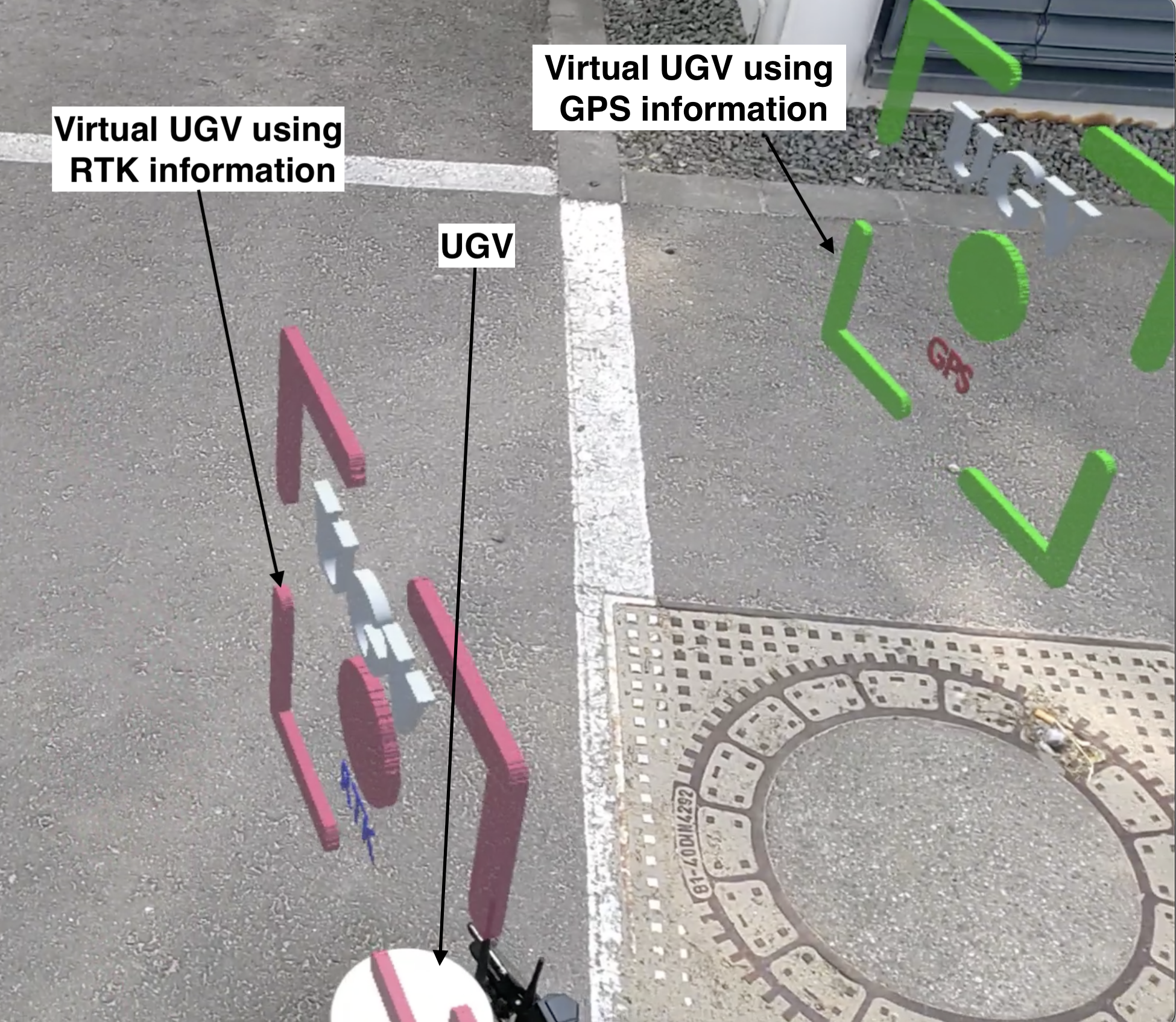}
     \caption{  Tracking a UGV using RTK  and GPS. Frame is captured from HoloLens v2. The virtual objects (rose and green rectangles) should be in the same position as the UGV such that the green rectangle represents the GPS value derived from the iPhone, and the rose rectangle represents the RTK value derived from RTK rover.}
        \label{RTKvsGPS}
\end{figure}

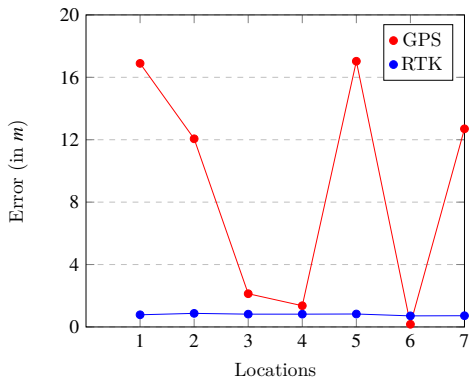
\begin{figure}
\centering
\resizebox{0.40\textwidth}{!}{  

\begin{tikzpicture}
\begin{axis}[
    xlabel={Locations},
    ylabel={Error (in $m$)},
    xmin=0, xmax=7,
    ymin=0, ymax=20,
    xtick={1,2,3,4,5,6,7},
    ytick={0,4,8,12,16,20},
    legend pos=north east, 
    ymajorgrids=true,
    grid style=dashed,
]

\addplot[
    only marks,
    mark=*,
    red,
    mark options={fill=red},
    nodes near coords,
    point meta=explicit symbolic,
    ] 
    coordinates {
    (1,16.893) 
    (2,12.06) 
    (3,2.13) 
    (4,1.36) 
    (5,17.03) 
    (6,0.17) 
    (7,12.7) 
    };

\addplot[
    only marks,
    mark=*,
    blue,
    mark options={fill=blue},
    nodes near coords,
    point meta=explicit symbolic,
    ] 
    coordinates {
    (1,0.78) 
    (2,0.87) 
    (3,0.82) 
    (4,0.82) 
    (5,0.83) 
    (6,0.71) 
    (7,0.72) 
    };
\addplot[red] coordinates {
  (1,16.893) 
    (2,12.06) 
    (3,2.13) 
    (4,1.36) 
    (5,17.03) 
    (6,0.17) 
    (7,12.7) 
};

\addplot[blue] coordinates {
    (1,0.78) 
    (2,0.87) 
    (3,0.82) 
    (4,0.82) 
    (5,0.83) 
     (6,0.71) 
    (7,0.72) 
};

\legend{GPS, RTK}
\end{axis}
\end{tikzpicture}}
\caption{Scatter plot illustrating the errors in distance measurements for several UGV locations obtained using both RTK and GPS systems. 
}
\label{fig_errors_distance_rtk_GPS}
\end{figure}

To accurately simulate a real-world scenario involving a CBRN incident in an urban environment, the system as illustrated in Figure \ref{system}  was thoroughly evaluated in a mixed use urban area containing a diverse array of structures such as multiple buildings, trees, and various other environmental elements that would typically surround the UGV. This complex and realistic urban landscape is illustrated in Figure \ref{fig:env}. One advantage of this system  is that it can still track objects even when they're not directly in sight, unlike systems that rely on images. As illustrated in Figure \ref{tracking_robot_car}, when the UGV is occluded by the tree, it can still be tracked via the AR application. This can be  useful, especially in situations where you need to see past obstacles. For example, in the context of a radiological incident and the management of CBRN, the utilization of OST-HMD by first responders could significantly enhance their situational awareness. These devices would enable them to gain a comprehensive understanding of the scenario, facilitating real-time observation of various elements, such as the location of the UGV. Notably, the latter could be equipped with systems similar to those integrated into our prototype  UGV, thereby extending the capabilities of the response team to effectively assess and mitigate the situation. \\ However, in this paper, we provide a preliminary evaluation of the system from the perspectives of functionality and accuracy. Further study is required to evaluate the system with end users, specifically first responders.\\
The experimental test begins by locating the RTK station and automatically determining its surveyed position using more than 1,000 positions of the SPP solution. This functionality is provided by the Swift Navigation Piksi Multi RTK GNSS system. Afterward, the UGV is positioned at a predetermined reference point. To calibrate the system, the person wearing the OST-HMD stands as close as possible to the UGV, holding a tablet to drive the UGV via a detected WiFi network. Both the person and the UGV face the north direction. Subsequently, the person wearing the OST-HMD turns on the HoloLens and runs the application. After the calibration step is complete, the person starts driving the UGV via the tablet and is free to move without any restrictions. A virtual object representing the UGV follows the physical UGV, visualized through the HoloLens. 
As mentioned in Section \ref{background}, in assessing GPS and RTK accuracy (see Figure \ref{RTKvsGPS}), two main methods are used: a) static: this involves comparing specific GPS/RTK location data with the real-world coordinates. It helps understand accuracy in fixed positions. b) dynamic: unlike static, dynamic analysis looks at the entire trajectory from GPS/RTK against real-world movement. This helps assess accuracy while in motion, useful for tasks like navigation and tracking. 
However, in our proposed methodology for system evaluation, applying the dynamic method has proven challenging. Synchronizing the movement of both the UGV and the HoloLens wearer to capture a full trajectory for error distance evaluation has presented difficulties. As a result, we propose a "semi-dynamic" approach, addressing \textit{RQ3: how well does the semi-dynamic approach adapt to the challenges of real-world UGV tracking evaluation compared to static and dynamic methods?} In this method, the UGV is driven and paused at various locations to measure error distances before completing the route for a more comprehensive evaluation. \\
To measure the error distance,  our approach (semi-dynamic) is grounded on the hypothesis that the distance between the physical camera position of the HoloLens and the virtual tracked object should ideally be zero when the wearer of the HoloLens stands in the same position as the UGV simultaneously  precisely in same position of RTK rover, disregarding any height difference. We conducted multiple iterations of this process, ensuring that error distances were recorded while standing as close as possible to the RTK rover antenna. 
Figure \ref{fig_errors_distance_rtk_GPS} illustrates the absolute error distances between the positions UGV measured using RTK information and GPS information derived from iPhone  and the HoloLens camera position, knowing that the y-axis of the HoloLens is set to zero, as our aim is to calculate the error in a 2D plane without considering height. 

As expected, we observe that the position of the virtual object using GPS values is jumping and unstable in densely populated urban areas. In our experiment,  the standard deviation for the different positions of a trajectory was approximately $7.453$ meters. Consequently, it's necessary to apply a filter to these data to reduce the jumping behavior. Conversely, RTK values exhibit a standard deviation of $0.126$ meters and provide more stable results.  As shown in the Figure \ref{fig_errors_distance_rtk_GPS}, a shifted offset value between the real object and the virtual one provided by RTK is approximately constant observed. In line with expectations, RTK offers greater precision than GPS, with an average of $0.745$ meters compared to $8.907$ meters provided by GPS.\\

Although it is well known that RTK offers significantly higher accuracy than standard GPS, the primary objective of this experimental test is to demonstrate that our methodology for the hybrid system of OST-HMD and RTK—specifically the processes of calibration and location updating—functions effectively, while maintaining the high accuracy provided by RTK.

\section{\uppercase{Limitations and future directions}}
To optimize the proposed OST-HMD and RTK system for outdoor tracking, we outline the current system's limitations and suggest future research directions addressing \textit{RQ4: What are the key research directions for optimizing the integration of OST-HMD with RTK systems for outdoor tracking?} These are categorized into three primary areas: 1) improving the accuracy of the tracking algorithm, 2) improving network latency, and 3) enhancing the visualization of tracked objects in OST-HMD. 
 \subsection{Accuracy}
 The accuracy of the proposed system could be significantly enhanced for real-world applications by taking the following points into account:
\begin{itemize}
    \item Accurate values for surveyed positions could increase the accuracy of the entire system. In our application we use an  automatic surveyed position which provides a relative  surveyed position. However, some countries provide an RTK station with surveyed position  that provides a correction position for a RTK rover which connect with it  via CORS network. This solution could save time during the calibration step because the setting up of station RTK doesn't be required. 
    \item The calibration step has a direct impact on the proposed system. Integrating other sensors, such as a compass, into OST-HMD could facilitate the orientation calibration process.
     \item  OST-HMDs are not recommended for outdoor deployment, as mentioned in the introduction. However, improving the quality of depth sensors to function outdoors under varying light conditions, which they currently handle well indoors, could directly improve the performance of vSLAM and enhance the adaptability of OST-HMDs for outdoor use. 
\end{itemize}
Hence, addressing these key areas—accurate surveyed positions, enhanced calibration processes with additional sensors, and improved depth sensor quality for outdoor use—can significantly enhance the effectiveness and adaptability of hybrid systems based on OST-HMDs and RTK systems in real-world applications.
\subsection {Network}
Improving network latency is critical for ensuring smooth and responsive communication. Currently, we set the refresh rate of data transmissions between the server and the UGV to 0.1 seconds. This rate is implemented to prevent the server from becoming overloaded by the high rate of incoming requests. However, it unfortunately results in a noticeable visual delay between the virtual representation of the UGV and its actual movements.
To overcome this challenge and enhance system performance, several key strategies can be implemented:
\begin{itemize}
    \item Concurrency with Multi-threading and Multiprocessing: Utilizing multi-threading and multiprocessing techniques enables the application to manage multiple connections simultaneously. By distributing workload across multiple threads or processes, the system can effectively handle a higher volume of requests without increasing latency.
    \item Network Optimization: Implementing Quality of Service (QoS) policies, optimizing network infrastructure, and ensuring direct and efficient routing paths between client devices and servers can further minimize latency. 
    \item Protocol Optimization: As described in Section \ref{server_application}, we chose socket connections designed to be efficient and minimize latency compared to traditional HTTPS requests. These connections provide full-duplex communication over a single TCP connection. However, based on our experimental tests and visual latency observations, we suggest considering UDP. This latter is a connectionless protocol that offers lower latency compared to TCP because it does not include the same level of error-checking and packet retransmission. However, it requires additional handling for reliability. 
    \item Predictive Rendering: Using predictive algorithms on the HoloLens to anticipate the UGV's movements based on its last known trajectory can compensate for short-term latency by rendering the UGV's position ahead of real-time updates.
\end{itemize}
Implementing these strategies could mitigate latency issues and deliver a seamless and user experience in our AR application.
\subsection{Visualisation}
 We recognize that a primary obstacle in implementing a suitable tracking system integrated with OST-HMD for outdoor environments, assuming we have accurate positioning derived from an RTK systems, is the perception of  alignment of virtual outdoor objects with their real counterparts. Work by \cite{pfeil2021distance} conducted a user study with 26 participants to help understand if distance perception is altered when viewing surroundings with a VST-HMD. Their results show that the inclusion of a ZED Mini pass-through camera causes a significant difference between normal, unrestricted viewing and that through a VST-HMD. However, there is less studies that tackle the perception through for OST-HMD for outdoor applications.
Hence, this task presents significant challenges, thus sparking numerous relevant research inquiries:
\begin{itemize}
    \item What is the optimal shape (2D form such rectangle, 3D form cube, etc.) to represent the virtual object that will  track and mimic the movements of the real-world object it follows?
    \item How does the perception of outdoor virtual objects change depending on their proximity to real objects? In a comparison of egocentric distance judgments with two different HMDs using indoor and outdoor virtual environments, study by \cite{creem2015egocentric} found greater underestimation in outdoor than indoor environments 
    \item What strategies can be employed to maintain alignment accuracy over varying distances and terrains? 
    \item   Formal comparison in the perception of virtual objects with and without using sun filter protection was not conducted in this paper. However, based on our experience, we recommend using sun filter protection for outdoor OST-HMD. Therefore, an interesting research question could be how environmental factors such as lighting conditions and weather impact the perceived alignment of virtual and real objects with and without using sun filter protection. Work by \cite{erickson2020exploring} investigated how environment lighting conditions impact the contrast of virtual imagery displayed to the user of an OST-HMD. Their results indicate that OST-HMD tends to lose nearly all contrast when taken into outdoor environments where illuminance levels are greater than 10,000 lux. Study by \cite{erickson2020exploring} recommends that beyond reducing environment light, future OST-HMD research should consider simulating and making adjustments to the intensive environment lighting in the outdoors.
\end{itemize}
Therefore, tackling these research requires—optimal shapes for virtual objects, perception changes based on object proximity, alignment strategies over varying terrains, effective occlusion management, and the impact of environmental factors on perceived alignment — could enhance the performance and usability of OST-HMD and RTK systems for outdoor tracking.


\section{\uppercase{Conclusion}}
Current OST-HMDs overcome many limitations of VST-HMDs but still face challenges when used in outdoor environments. This paper proposes an approach to integrate data from an RTK system and track this information using the vSLAM algorithm in OST-HMD. We hypothesize that combining UGV, OST-HMD, and accurate positioning can enhance the ability of first responders to manage incidents, particularly by improving their capability to visualize occluded information, thereby increasing situational awareness and safety.

Our system consists of three core components: (1) a web server that receives data from a UGV and transmits it to OST-HMD via a socket connection; (2) a UGV equipped with an RTK rover system; and (3) the HoloLens 2, serving as the OST-HMD. A detailed calibration step, which ensures accurate global tracking of the user's position and orientation, is illustrated.

In this paper, we present a preliminary evaluation of the system in terms of functionality and accuracy. Further research is necessary to assess the system with end users, specifically first responders. Additionally, the paper raises critical research questions for further development in the field, identifying three key areas (accuracy, networking, and visualization) comprising 10 specific topics.

In conclusion, this paper advances the state of the art in outdoor RTK positioning with OST-HMD, proposing a comprehensive system for visualizing UGV data via OST-HMD while also highlighting areas for future research.
\section*{\uppercase{Acknowledgements}}
Muhannad Ismael and Maël Cornil are supported by the Luxembourg Institute of Science and Technology (LIST) and the Luxembourg National Research Fund (FNR) under the RISARX project (Grant Number 15340411). We express our gratitude to Dr. Roderick McCall for his support and insightful discussions, Mr. Christian Moll and Mr. Johannes Hermen for their assistance in constructing the UGV, and Dr. Mohamed Saifeddine Hadj Sassi for his logistical coordination during the outdoor evaluation. Moreover, this paper has been accepted as a short paper in VISIGRAPP\footnote{\url{https://www.scitepress.org/Papers/2025/131326/131326.pdf}}.
\bibliographystyle{apalike}

{\small

\begin{thebibliography}{}

\bibitem[Arth et~al., 2015]{arth2015instant}
Arth, C., Pirchheim, C., Ventura, J., Schmalstieg, D., and Lepetit, V. (2015).
\newblock Instant outdoor localization and slam initialization from 2.5 d maps.
\newblock {\em IEEE Transactions on Visualization \& Computer Graphics}, 21(11):1309--1318.

\bibitem[Behzadan and Kamat, 2005]{behzadan2005visualization}
Behzadan, A.~H. and Kamat, V.~R. (2005).
\newblock Visualization of construction graphics in outdoor augmented reality.
\newblock In {\em Proceedings of the Winter Simulation Conference, 2005.}, pages 7--pp. IEEE.

\bibitem[Benjumea et~al., 2021]{DBLP}
Benjumea, A., Teeti, I., Cuzzolin, F., and Bradley, A. (2021).
\newblock {YOLO-Z:} improving small object detection in yolov5 for autonomous vehicles.
\newblock {\em CoRR}, abs/2112.11798.

\bibitem[Burkard and Fuchs-Kittowski, 2020]{burkard2020user}
Burkard, S. and Fuchs-Kittowski, F. (2020).
\newblock User-aided global registration method using geospatial 3d data for large-scale mobile outdoor augmented reality.
\newblock In {\em 2020 IEEE International Symposium on Mixed and Augmented Reality Adjunct (ISMAR-Adjunct)}, pages 104--109. IEEE.

\bibitem[Creem-Regehr et~al., 2015]{creem2015egocentric}
Creem-Regehr, S.~H., Stefanucci, J.~K., Thompson, W.~B., Nash, N., and McCardell, M. (2015).
\newblock Egocentric distance perception in the oculus rift (dk2).
\newblock In {\em Proceedings of the ACM SIGGRAPH symposium on applied perception}, pages 47--50.

\bibitem[Dahne and Karigiannis, 2002]{dahne2002archeoguide}
Dahne, P. and Karigiannis, J.~N. (2002).
\newblock Archeoguide: System architecture of a mobile outdoor augmented reality system.
\newblock In {\em Proceedings. International Symposium on Mixed and Augmented Reality}, pages 263--264. IEEE.

\bibitem[De~Pace and Kaufmann, 2023]{de2023systematic}
De~Pace, F. and Kaufmann, H. (2023).
\newblock A systematic evaluation of an rtk-gps device for wearable augmented reality.
\newblock {\em Virtual Reality}, 27(4):3165--3179.

\bibitem[Erickson et~al., 2020]{erickson2020exploring}
Erickson, A., Kim, K., Bruder, G., and Welch, G.~F. (2020).
\newblock Exploring the limitations of environment lighting on optical see-through head-mounted displays.
\newblock In {\em Proceedings of the 2020 ACM Symposium on Spatial User Interaction}, pages 1--8.

\bibitem[Gal et~al., 2001]{gal2001cartogam}
Gal, O., Izac, C., Jean, F., Lain{\'e}, F., L{\'e}v{\^e}que, C., and Nguyen, A. (2001).
\newblock Cartogam--a portable gamma camera for remote localisation of radioactive sources in nuclear facilities.
\newblock {\em Nuclear Instruments and Methods in Physics Research Section A: Accelerators, Spectrometers, Detectors and Associated Equipment}, 460(1):138--145.

\bibitem[Gan-Mor et~al., 2007]{gan2007implement}
Gan-Mor, S., Clark, R.~L., and Upchurch, B.~L. (2007).
\newblock Implement lateral position accuracy under rtk-gps tractor guidance.
\newblock {\em Computers and Electronics in Agriculture}, 59(1-2):31--38.

\bibitem[H{\"o}llerer et~al., 1999]{hollerer1999exploring}
H{\"o}llerer, T., Feiner, S., Terauchi, T., Rashid, G., and Hallaway, D. (1999).
\newblock Exploring mars: developing indoor and outdoor user interfaces to a mobile augmented reality system.
\newblock {\em Computers \& Graphics}, 23(6):779--785.

\bibitem[Klein and Murray, 2007]{klein07parallel}
Klein, G. and Murray, D. (2007).
\newblock Parallel tracking and mapping for small {AR} workspaces.
\newblock In {\em Proc. Sixth {IEEE} and {ACM} International Symposium on Mixed and Augmented Reality {(ISMAR'07)}}, Nara, Japan.

\bibitem[Kluga et~al., 2014]{kluga2014state}
Kluga, A., Mitrofanovs, I., Kluga, J., and Jeralovics, V. (2014).
\newblock State and dynamic precision research using two gps receivers with rtk.
\newblock In {\em 2014 14th Biennial Baltic Electronic Conference (BEC)}, pages 141--144. IEEE.

\bibitem[Lim et~al., 2012]{lim2012real}
Lim, H., Sinha, S.~N., Cohen, M.~F., and Uyttendaele, M. (2012).
\newblock Real-time image-based 6-dof localization in large-scale environments.
\newblock In {\em 2012 IEEE conference on computer vision and pattern recognition}, pages 1043--1050. IEEE.

\bibitem[Ling et~al., 2019]{8798315}
Ling, F.~F., Elvezio, C., Bullock, J., Henderson, S., and Feiner, S. (2019).
\newblock A hybrid rtk gnss and slam outdoor augmented reality system.
\newblock In {\em 2019 IEEE Conference on Virtual Reality and 3D User Interfaces (VR)}, pages 1044--1045.

\bibitem[Morales and Tsubouchi, 2007]{morales2007dgps}
Morales, Y. and Tsubouchi, T. (2007).
\newblock Dgps, rtk-gps and starfire dgps performance under tree shading environments.
\newblock In {\em 2007 IEEE International Conference on Integration Technology}, pages 519--524. IEEE.

\bibitem[Niu et~al., 2020]{niu2020continuous}
Niu, Z., Zhao, X., Sun, J., Tao, L., and Zhu, B. (2020).
\newblock A continuous positioning algorithm based on rtk and vi-slam with smartphones.
\newblock {\em IEEE Access}, 8:185638--185650.

\bibitem[Oskiper et~al., 2012]{oskiper2012multi}
Oskiper, T., Samarasekera, S., and Kumar, R. (2012).
\newblock Multi-sensor navigation algorithm using monocular camera, imu and gps for large scale augmented reality.
\newblock In {\em 2012 IEEE international symposium on mixed and augmented reality (ISMAR)}, pages 71--80. IEEE.

\bibitem[Pfeil et~al., 2021]{pfeil2021distance}
Pfeil, K., Masnadi, S., Belga, J., Sera-Josef, J.-V.~T., and LaViola, J. (2021).
\newblock Distance perception with a video see-through head-mounted display.
\newblock In {\em Proceedings of the 2021 CHI Conference on Human Factors in Computing Systems}, pages 1--9.

\bibitem[Rao et~al., 2017]{rao2017mobile}
Rao, J., Qiao, Y., Ren, F., Wang, J., and Du, Q. (2017).
\newblock A mobile outdoor augmented reality method combining deep learning object detection and spatial relationships for geovisualization.
\newblock {\em Sensors}, 17(9):1951.

\bibitem[Roberts et~al., 2002]{roberts2002use}
Roberts, G.~W., Evans, A., Dodson, A., Denby, B., Cooper, S., Hollands, R., et~al. (2002).
\newblock The use of augmented reality, gps and ins for subsurface data visualization.
\newblock In {\em FIG XXII international congress}, volume~4, pages 1--12.

\bibitem[Safrel et~al., 2018]{safrel2018accuracy}
Safrel, I., Julianto, E.~N., and Usman, N.~Q. (2018).
\newblock Accuracy comparison between gps real time kinematic (rtk) method and total station to determine the coordinate of an area.
\newblock {\em Jurnal Teknik Sipil Dan Perencanaan}, 20(2):123--130.

\bibitem[Satheesan, 2024]{satheesan2024real}
Satheesan, A. (2024).
\newblock Real-time augmented reality based operator assistance for driving cut-to-length forest machines.

\bibitem[Schall et~al., 2009]{schall2009global}
Schall, G., Wagner, D., Reitmayr, G., Taichmann, E., Wieser, M., Schmalstieg, D., and Hofmann-Wellenhof, B. (2009).
\newblock Global pose estimation using multi-sensor fusion for outdoor augmented reality.
\newblock In {\em 2009 8th ieee international symposium on mixed and augmented reality}, pages 153--162. IEEE.

\bibitem[Singh et~al., 2022]{singh2022augmented}
Singh, S., Singh, J., Shah, B., Sehra, S.~S., and Ali, F. (2022).
\newblock Augmented reality and gps-based resource efficient navigation system for outdoor environments: Integrating device camera, sensors, and storage.
\newblock {\em Sustainability}, 14(19):12720.

\bibitem[Stranner et~al., 2019]{stranner2019high}
Stranner, M., Arth, C., Schmalstieg, D., and Fleck, P. (2019).
\newblock A high-precision localization device for outdoor augmented reality.
\newblock In {\em 2019 IEEE International Symposium on Mixed and Augmented Reality Adjunct (ISMAR-Adjunct)}, pages 37--41. IEEE.

\bibitem[Swan et~al., 2003]{Swan2003PerceptualAE}
Swan, J.~E., Hix, D., and Gabbard, J.~L. (2003).
\newblock Perceptual and ergonomic issues in mobile augmented reality for urban operations.

\bibitem[Tomaszewski et~al., 2020]{tomaszewski2020assessment}
Tomaszewski, D., Wielgosz, P., Rapi{\'n}ski, J., Krypiak-Gregorczyk, A., Ka{\'z}mierczak, R., Hern{\'a}ndez-Pajares, M., Yang, H., and Or{\'u}sP{\'e}rez, R. (2020).
\newblock Assessment of centre national d’{\`e}tudes spatiales real-time ionosphere maps in instantaneous precise real-time kinematic positioning over medium and long baselines.
\newblock {\em Sensors}, 20(8):2293.

\bibitem[Ungureanu et~al., 2020]{ungureanu2020hololens}
Ungureanu, D., Bogo, F., Galliani, S., Sama, P., Duan, X., Meekhof, C., St{\"u}hmer, J., Cashman, T.~J., Tekin, B., Sch{\"o}nberger, J.~L., et~al. (2020).
\newblock Hololens 2 research mode as a tool for computer vision research.
\newblock {\em arXiv preprint arXiv:2008.11239}.

\bibitem[Wi{\'s}niewski et~al., 2013]{wisniewski2013evaluation}
Wi{\'s}niewski, B., Bruniecki, K., and Moszy{\'n}ski, M. (2013).
\newblock Evaluation of rtklib's positioning accuracy usingn low-cost gnss receiver and asg-eupos.
\newblock {\em TransNav: International Journal on Marine Navigation and Safety of Sea Transportation}, 7(1):79--85.

\bibitem[Zari et~al., 2023]{zari2023magic}
Zari, G., Condino, S., Cutolo, F., and Ferrari, V. (2023).
\newblock Magic leap 1 versus microsoft hololens 2 for the visualization of 3d content obtained from radiological images.
\newblock {\em Sensors}, 23(6):3040.

\bibitem[Zhang and Kosecka, 2006]{zhang2006image}
Zhang, W. and Kosecka, J. (2006).
\newblock Image based localization in urban environments.
\newblock In {\em Third international symposium on 3D data processing, visualization, and transmission (3DPVT'06)}, pages 33--40. IEEE.

\end{thebibliography}


}



\end{document}